\documentclass[11pt]{article}%
\usepackage[dvips]{graphicx}
\usepackage{graphicx}
\usepackage{wrapfig}
\usepackage{color}
\usepackage{boxedminipage}
\usepackage{amsfonts}
\usepackage{amsmath,amssymb}
\usepackage{url}
\usepackage{helvet} 
\usepackage{cite}
\usepackage{soul}
\sethlcolor{green}

\makeatletter 
\renewcommand{\section}{\@startsection{section}{1}{0mm}{2mm}{2mm}{\Large\normalfont\bfseries}}
\renewcommand{\subsection}{\@startsection{subsection}{2}{0mm}{1mm}{1mm}{\large\normalfont\bfseries}}
\renewcommand{\subsubsection}{\@startsection{subsubsection}{3}{0mm}{1mm}{0.5mm}{\normalfont\bfseries}}
\makeatother 

\newcommand{\eq}[1]{Eq.\ (\ref{#1})}

\newcommand{\flux}{\mathrm{Flux}}

\newcommand{\kmunu}{k^{\mu \nu}}

\newcommand{\mfpt}{\mathrm{MFPT}}

\newcommand{\pab}{p^{\alpha}}

\newcommand{\pba}{p^{\beta}}

\newcommand{\weexact}{Zhang2010a}
\newcommand{\wess}{Bhatt2010a}

\begin{document}
\bibliographystyle{unsrt}
\title{Simultaneous computation of dynamical and equilibrium information using a weighted ensemble of trajectories}
\author{Ernesto Su{\'a}rez$^{1\dagger}$, Steven Lettieri$^{1\dagger}$, Matthew C.\ Zwier$^2$, Carsen A.\ Stringer$^3$,\\ Sundar Raman Subramanian$^1$, Lillian T.\ Chong$^2$,
and Daniel M.\ Zuckerman$^{1*}$\\
\\
1. Department of Computational and Systems Biology, University of Pittsburgh\\
2. Department of Chemistry, University of Pittsburgh\\
3. Gatsby Computational Neuroscience Unit, University College London\\
\\
$\dagger$ Equal contributions\\
* Corresponding author: ddmmzz@pitt.edu}

\maketitle

\section*{Abstract}
Equilibrium formally can be represented as an ensemble of uncoupled systems undergoing unbiased dynamics in which detailed balance is maintained. Many non-equilibrium processes can be described by suitable subsets of the equilibrium ensemble.  Here, we employ the ``weighted ensemble'' (WE) simulation protocol [Huber and Kim, \emph{Biophys. J.}, 1996] to generate equilibrium trajectory ensembles and extract non-equilibrium subsets for computing kinetic quantities.  States do not need to be chosen in advance.  The procedure formally allows estimation of kinetic rates between arbitrary states chosen after the simulation, along with their equilibrium populations.  We also describe a related history-dependent matrix procedure for estimating equilibrium and non-equilibrium observables when phase space has been divided into arbitrary non-Markovian regions, whether in WE or ordinary simulation.  
In this proof-of-principle study, these methods are successfully applied and validated on two molecular systems: explicitly solvated methane association and the implicitly solvated Ala4 peptide.  We comment on challenges remaining in WE calculations.

\section{Introduction}
Although it is textbook knowledge that the functions of biomacromolecules are strongly coupled to their conformational motions and fluctuations \cite{Berg-2002}, computer simulation of such motions has been a challenge for decades \cite{Zuckerman2011}.
Typically, distinct algorithms are employed to estimate equilibrium quantities (e.g., \cite{Swendsen-1986,Zheng2008a}) and dynamical properties (e.g., \cite{Huber-1996,Bolhuis2002,Allen2005,Warmflash2007,Vanden-Eijnden2009, Elber2004}).
In principle, a single long dynamics trajectory would be sufficient to determine both equilibrium and dynamical properties \cite{Shaw2010}, but such simulations remain impractical for most systems of interest.

Aside from straightforward simulations, more technical approaches that can yield both equilibrium and dynamical simulation, sometimes under minor assumptions, have drawn increasing attention. Replica exchange molecular dynamics (REMD)\cite{Sugita1999,Buchete2008}, is a common method to improve the conformational sampling, where independent copies of the system are simulated in parallel, each at different temperature. Periodically, the algorithm attempts to exchange replicas using a Monte Carlo procedure. With this strategy, is possible in principle to extract kinetic information from continuous trajectory segments between replica exchanges \cite{Buchete2008}. In this case the authors took advantage of a Markov model to obtain converged estimates. 

The adaptive seeding method (ASM) \cite{Huang2009} is fairly similar. The phase space is explored by REMD or any other of the so-called generalized ensemble (GE) algorithms where random walks are done in temperature space. Then a Markov state model (MSM) is built to identify all the metastable states. New constant temperature simulations are done at the temperature of interest from each metastable state in a process called seeding. Finally, the MSM is used to extract the correct equilibrium populations from the seeding simulations. Markov models have also been used by Noe et al. \cite{Noe2009} in combination with short, off-equilibrium simulations to construct the the equilibrium ensemble of folding pathways of a protein. 

Milestoning is another strategy where is possible to perform both equilibrium and rate calculations. In this approach, the system is partitioned into cells by dividing hypersurfaces (Milestones) and transitions are computed between nearby hypersurfaces. The observables are obtained from the statistics of these transitions \cite{Faradjian2004,West2007}. However, this method assumes a complete loss of memory at each interface. 

Moroni et al have developed a related method, transition interface sampling TIS, which is based on the computation of crossing probabilities of a set of interfaces between the initial and final states \cite{Erp2003,Moroni2004}. A variant of this method, partial path TIS (PPTIS), has been used to compute free-energy barriers as well as rates constants from a single calculation \cite{Moroni2005}. This method introduces a stronger history dependence than Milestoning, however assumes a loss of time correlations in the transition paths over a distance of two interfaces.

Unlike TIS/PPTIS or Milestoning, forward flux sampling (FFS) is not limited to systems in equilibrium. Like in those methods, a series of interfaces are used to compute the rate constant. Nevertheless, FFS does not make the Markovian assumption that the distribution of paths at the interfaces is independent on the path histories.

\begin{figure}\label{fig:equil}
\begin{center}
\hspace*{-0.3in}
\includegraphics[width=0.9\textwidth]{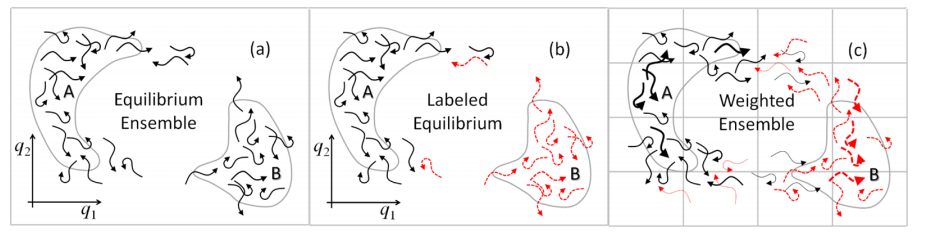}
\end{center}
\caption {
\footnotesize
Equilibrium in different representations.
(a) Ensemble of trajectories with arrow tips indicating the instantaneous configuration and tails showing recent history in the space of two schematic coordinates $q_1$ and $q_2$.
States A and B, shown in grey, are two arbitrary regions of phase space.
(b) Dissection into two subsets based on whether a trajectory was most recently in state A (black solid arrows, the ``$\alpha$'' steady state) or state B (red dashed, the ``$\beta$'' steady state).
(c) Statistically equivalent ensemble of weighted trajectories, with arrow thickness suggesting weight.
Configuration space has been divided into cells (``bins'') which each contain an equal number of trajectories.
}
\end{figure}

The ``weighted ensemble'' (WE) simulation strategy \cite{Huber-1996} (see Fig. \ref{fig:equil}), which has a rigorous basis as a path-sampling method \cite{Zhang2010a}, has also been suggested as an approach for computation of both equilibrium and non-equilibrium properties \cite{Bhatt2010a,Bhatt2012}.  Although WE was originally developed as a tool for characterizing non-equilibrium dynamical pathways and rates (e.g., \cite{Huber-1996, Rojnuckarin1998, Zhang2007, Bhatt2009a,Zwier2011}), the strategy was extended to steady-state conditions including equilibrium \cite{Bhatt2010a}.  The simultaneous computation of equilibrium and kinetic properties using WE was demonstrated with configuration space separated into two states by a dividing surface \cite{Bhatt2012} and later for arbitrary states defined in advance of a simulation \cite{Darve-chapter}.

Here, we further develop the capability of WE simulation to calculate equilibrium and non-equilibrium quantities simultaneously in several ways that may be important for future studies of increasingly complex systems. (i) The approach described below permits the calculation of rates between arbitrary states, which can be defined \emph{after} a simulation has been completed.  In a complex system, the most important physical states, including intermediates, generally will not be obvious prior to simulation.  Further, the present approach opens up the possibility to use rate calculations to aid in the state-definition process.  (ii) The non-Markovian analysis described here enables unbiased rate calculations in the typical case where ``bins'' used by WE simulation do not exhibit Markovian behavior.  The analysis is general and can be applied outside the WE context, including the analysis of ordinary long trajectories.  (iii) The non-Markovian analysis can improve the efficiency of WE simulations by yielding accurate estimates of observables from shorter simulations.  The analysis is based on a previously suggested decomposition of the equilibrium ensemble into two non-equilibrium steady states \cite{Vanden-Eijnden2009, Dickson2009a, Moroni2005, Valeriani2007,Bhatt2011}.

Generally speaking, WE provides an attractive basis for complex simulations.  WE is easily parallelizable because it employs multiple trajectories, and was recently used with 3,500 cores \cite{ZwierSubmitted}.  WE algorithms lend themselves to a scripting-like implementation which has been employed to study a wide range of stochastic systems via regular molecular dynamics \cite{Zwier2011}, Monte Carlo \cite{Zhang2007}, the string strategy \cite{Aldelman2013}, and Gillespie-algorithm dynamics of chemical kinetic networks \cite{Donovan2013}.

\section{Theoretical formulation}

WE simulation uses multiple simultaneous trajectories, with weights that sum to one, that are occasionally coupled by replication or combination events every $\tau$ units of time \cite{Huber-1996}.
The coupling events typically are governed by a static partition of configuration space into ``bins'' (Fig.\ \ref{fig:equil}c), although dynamical/adaptive bins may be used \cite{\weexact}.
In the case of static bins, when one or more trajectories enters an unoccupied bin, those trajectories are replicated so that their count conforms to a (typically) preset value, $M$. Replicated ``daughter'' trajectories inherit equal shares of the parent's weight. If more than $M$ trajectories are found to occupy a bin, trajectories are combined statistically in a pairwise fashion until $M$ remain with weight from pruned trajectories assigned to others in the same bin. These procedures are carried out in such a way that dynamics remain statistically unbiased \cite{\weexact}.
This study does not adjust weights according to previously developed reweighting procedures \cite{\wess} during the simulation.  Rather, the WE simulations described here are long enough to permit relaxation to the equilibrium state.

\subsection{Direct calculation of observables}
Once the equilibrium state is reached in a WE simulation, meaning that there is a detailed balance of probability flow between any two states, equilibrium observables such as state populations or a potential of mean force can be calculated simply by summing trajectory weights in the corresponding regions of phase space.  We term this ``direct'' estimation of observables.

To calculate rates, the equilibrium set of trajectories (Fig~\ref{fig:equil}a) is decomposed into two steady states as shown in Fig. \ref{fig:equil}b: the $\alpha$ steady state consisting of trajectories more recently in A than B, and the $\beta$ steady state with those most recently in B \cite{Vanden-Eijnden2009, Bhatt2011}; these were denoted ``AB'' and ``BA'' steady states, respectively, in Ref. \cite{Bhatt2011}. Trajectories are ``labeled'' according to the last state visited, i.e., classified as $\alpha$ or $\beta$, during a WE simulation or in a post-simulation analysis (``post-analysis '').
The \emph{direct} rate $k_{AB}$ estimate is computed from the probability arriving to the final state \cite{Vanden-Eijnden2009,Bhatt2010a, Zuckerman2010,Zheng2008a,Allen2005,Moroni2005} via

\begin{equation}
\label{rate}
k_{AB}=\frac{1}{\mfpt(A \rightarrow B)}
  = \frac{ \flux(A \rightarrow B | \alpha) } { p(\alpha) } \; ,
\end{equation}
where MFPT is the mean-first-passage time, $\flux(A \rightarrow B | \alpha) $
 is the probability per unit time arriving to state B in the $\alpha$ steady state and $p(\alpha)$ is the total probability in the $\alpha$ steady state. By construction $p(\alpha)+p(\beta)=1$. Normalizing by $p(\alpha)$ effectively excludes the reverse steady state and the rate calculation only ``sees'' the uni-directional $\alpha$ steady state as in Ref. \cite{Bhatt2010a}. An expression analogous to \eq{rate} applies for $k_{BA}$. 
Also note that the effective first order rate constant, defined by $\flux(A \rightarrow B | \alpha) /p^{\mathrm{eq}}_A$, can be determined from equilibrium WE simulation because $p^{\mathrm{eq}}_A$ can be directly computed by summing weights in A.

We note that analogous direct calculation of observables can be performed from an equilibrium ensemble of unweighted (i.e., ``brute force'') trajectories by assigning equal weights to each.

\subsection{Non-Markovian matrix calculation of observables}
Beyond the direct estimates of observables based on trajectory weights, we also generalize previous matrix formulations for non-equilibrium steady states \cite{Dickson2009a,Dickson2011,Vanden-Eijnden2009} into an equilibrium formulation that explicitly accounts for the embedded steady states (as in Fig. 1b,c).  These non-Markovian matrix estimates are tested below and may prove important for future WE studies using shorter simulations, as described in the Discussion.

\begin{figure}
\begin{center}
\includegraphics[width=0.9\textwidth]{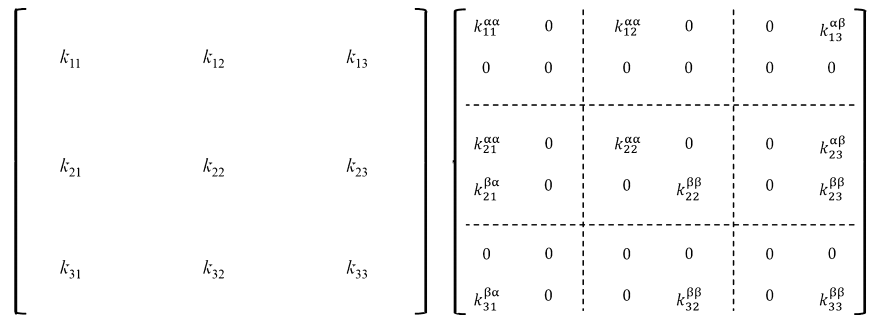}
\end{center}
\caption {
\footnotesize
\label{fig:matrices}
Constructing a labeled rate matrix for unbiased calculations.
For purposes of illustration, here state A consists solely of bin 1 and state B solely of bin 3.
Left: A traditional rate matrix with history-blind elements. The rate $k_{ij}$ gives the conditional probability for transitioning from bin $i$ to bin $j$ in a fixed time increment, regardless of previous history.
Right: The labeled rate matrix accounting for history.  The element $k_{ij}^{\mu \nu}$ is the conditional probability for the $i$ to $j$ transition for trajectories initially in the $\mu$ sub-ensemble which transition to the $\nu$ sub-ensemble, where $\mu$ and $\nu$ are either $\alpha$ or $\beta$.
The labeled rate matrix correctly assigns the $\alpha$ and $\beta$ sub-populations of each bin, whereas the traditional matrix may not.
}
\end{figure}

Our matrix approach explicitly uses the decomposition of the equilibrium population into $\alpha$ and $\beta$ components for each bin $i$: 
\begin{equation}
\label{colorpartition}
 p^{\mathrm{eq}}_i = p^{\alpha}_i + p^{\beta}_i
\end{equation}
which implies $p(\alpha) = \sum_i{p^{\alpha}_i}$ and $p(\beta) = \sum_i{p^{\beta}_i}$.  We called this a ``labeled'' analysis.   
Thus, with $N$ bins, a set of $2N$ probabilities is required rather than $N$.  Similarly, a $2N \times 2N$ rate matrix is required: $k^{\mu\nu}_{ij}$, where $\mu$ and $\nu$ can be either the $\alpha$ or $\beta$ subsets of trajectories.  See Fig. \ref{fig:matrices}.  Each of the previously considered $k_{ij}$ rate elements is thus decomposed into four history-dependent elements which account for whether the particular trajectory was last in state A or B and whether the trajectory transitions between the $\alpha$ and $\beta$ subsets. The analysis assumes states consist strictly of one or more bins, but this is always possible in a post-analysis without loss of generality. In other words, given the flexibility we have when we define the bins, is not a real limitation that the states have to be strictly constituted by bins.

We wish to emphasize that this analysis is ``non-Markovian'' because we are explicitly including history information (i.e., $\alpha$ and $\beta$ labels) in the new $2N \times 2N$ rate matrix. Once the matrix is built, the steady state observables are obtained using the same mathematical formalism that would be used in a regular Markov model. However, the matrix should be seen as a tool of linear algebra and not as embodying any physical assumptions.

Note that more than half the $k^{\mu\nu}_{ij}$ elements are zero. For example, consider a bin in the `intermediate' region (neither A nor B), such as bin 2 in Fig. \ref{fig:matrices}. 
In this region an $\alpha$ trajectory cannot change into a $\beta$ trajectory, nor vice versa; hence rates for these processes are zero. Similarly, an $\alpha$ trajectory in the intermediate region which enters a bin in B must turn into a $\beta$ trajectory, so the rate will always be zero to the $\alpha$ components of bins in B.

The non-Markovian results below stem from the division into $\alpha$ and $\beta$ steady states, but several steps are required.

First, rates among bins are estimated in a post-analysis as
\begin{equation}
\label{secondaverage}
k^{\mu\nu}_{ij}=\frac{\langle \omega^{\mu\nu}_{ij} \rangle_2}{\langle \omega^{\mu}_i \rangle},
\end{equation}
where $\omega^{\mu\nu}_{ij}$ is the flux, for a given iteration, from bin $i$ to $j$ of trajectories only with initial and final `labels' $\mu$ and $\nu$ respectively, while $\omega^{\mu}_i$ is the population labeled as $\mu$ which is initially in $i$. 
The subscript "2" in the numerator indicates that the rate $k^{\mu\nu}_{ij}$ is estimated to be non-zero only when more than one transition is observed; after the second event, all events are included, from the first one, to avoid bias. 
The requirement for two transitions was found to greatly enhance numerical stability in estimating fluxes and rates between macroscopic states: rates estimated from single events exhibit large fluctuations.

Notice that Eq.~\ref{secondaverage} is a \emph{ratio of averages} and differs from the average ratio $\langle \omega^{\mu\nu}_{ij} / \omega^{\mu}_i \rangle$, which might seem equally or more ``natural.''
However, our data below show that only Eq.\ \ref{secondaverage} yields unbiased estimates. 
The difference between the two estimators indicates that transitions are correlated with trajectory weights.  
Perhaps more importantly, the average ratio places less importance on high weight transitions occurring -- due to the ``instantaneous'' normalization -- and so, in a time-averaging sense, may be incorrect.  
That is, low-weight transitions count as heavily as high-weight events, which evidently biases the rate estimate.  
In the ratio of averages, high-weight events appropriately count more.

To obtain ``macroscopic'' rates between states consisting of arbitrary sets of bins (noting that arbitrary bins can be employed in a post-analysis), we calculate ``labeled'' fluxes for use in Eq.\ \ref{rate} via
\begin{equation}
\label{labeled-flux}
\flux(A \rightarrow B | \alpha) = \sum_{i,j} \, \pab_i k^{\alpha \beta}_{ij}
\hspace{1cm}
\flux(B \rightarrow A | \beta) = \sum_{i,j} \, \pba_i k^{\beta \alpha}_{ij} \; .
\end{equation}
The labeled bin populations $\pab_i$ and $\pba_i$ are obtained from the steady-state solution of the labeled rate matrix $K=\{\kmunu_{ij}\}$.

A summary of the ``labeled'' or non-Markovian matrix procedure for estimating rates between arbitrary states is as follows.
First, we obtain the labeled rate matrix $K=\{\kmunu_{ij}\}$ using Eq.\ \ref{secondaverage} to average inter-bin transitions. 
Second, we solve the matrix problem $K^T\textbf{p}_{SS}=\textbf{p}_{SS}$, yielding the steady state solution $\textbf{p}_{SS}$. 
Notice that the \emph{equilibrium} bin populations can be computed by Eq.~\ref{colorpartition}. 
Then, the steady state solution $\textbf{p}_{SS}$ along with the labeled rate matrix elements are used to calculate the $\alpha$ flux entering state B and the $\beta$ flux entering A. (Eq.~\ref{labeled-flux}). 
Finally, the MFPT values are obtained from Eq.~\ref{rate}.
In the graphs below, each non-Markovian estimate shown is from the matrix solution using the $\kmunu_{ij}$ rates calculated based on all data obtained until the given iteration of the simulation.

The non-Markovian matrix formulation exhibits a number of desirable properties: (i) Unlike with unlabeled (i.e., implicitly Markovian) analysis, kinetic properties will be unbiased as shown below. (ii) Solution of both the $\alpha$ and $\beta$ steady states is performed simultaneously via a standard Markov-state-like analysis of the $k^{\mu\nu}_{ij}$ rate matrix. By contrast, if the $\alpha$ and $\beta$ steady states are independently solved within a Markov formalism, there can be substantial ambiguity in how to assign feedback from the target to initial state when the initial state consists of more than one bin. (iii) The labeled formulation guarantees, by construction, the flux balance intrinsic to equilibrium, namely, $\flux(A \rightarrow B | \alpha) =\flux(B \rightarrow A | \beta)$. 
(iv) The analysis can be performed using arbitrary bins (and states defined as sets of these bins). It is not necessary to employ the bins originally used to run the WE simulation because a post-analysis can calculate rates among any regions of configuration space.  (v) The analysis is equally applicable to ordinary brute-force simulations.

\subsection{Markovian matrix calculation of observables}
\label{sec:nocolorestimation}
For reference, we also perform a traditional Markov analysis of the trajectories, which will prove to yield biased rate estimates because most divisions of configuration space (e.g., WE bins) are not true Markovian states. 

The Markov analysis proceeds \emph{without} labeling the trajectories. 
Elements of the rate matrix are estimated as
\begin{equation}
\label{regularrates}
k_{ij}=\langle w_{ij}\rangle_2/\langle w_i \rangle,
\end{equation}
where the subscript ``2'' again means that we only estimate a rate as non-zero once at least two transitions from $i$ to $j$ have occurred. 
Bin populations are then computed by solving for the steady-state solution of the Markov matrix with elements $k_{ij}$. 

The computation of an MFPT requires the use of source (A) and sink (B) states.
This task is automatically performed within the labeled formalism previously described.
Hence, we determine Markovian macroscopic rates by substituting the Markovian $k_{ij}$ for all non-zero elements of the $k^{\mu\nu}_{ij}$.
We emphasize that this is merely an accounting trick to establish sources and sinks and simultaneously measure both A-to-B and B-to-A fluxes/rates.

We perform a smoothing operation on the macroscopic Markovian rates because otherwise the data are fairly noisy.
The MFPT results shown for the Markovian matrix analysis are running averages based on the last 50\% of the estimates (where each estimate is from the matrix solution using $k_{ij}$ estimates from all data obtained until the particular iteration).
We confirmed numerically that such smoothing did not contribute bias to any of the MFPT estimates.

\section{Model systems and simulation details}
Weighted ensemble simulations were performed on two systems: the alanine tetrapeptide (Ala4) solvated implicitly and a pair of explicitly solvated methane molecules. 
All simulations were performed at 300K with stochastic thermostat. Friction constants of 5.0 and 1.0 ps$^{-1}$ were used for Ala4 and methane systems respectively. The molecular dynamics time step used for all systems was $\Delta t$ = 2 fs. An iteration is defined to be the simultaneous propagation of all trajectories in the ensemble for some amount of time, $\tau$. 
In these studies, a value of $\tau = 2500 \Delta t$ is used for Ala4 and $\tau = 250 \Delta t$ for the methane-methane system.

For Ala4, the all-atom AMBER ff99SB	 force-field \cite{ff99SB} with implicit GB/SA solvent and no cutoff for the evaluation of non-bonded interactions was simulated using the AMBER 11 software package \cite{amber11}. The Hawkins, Cramer, Truhlar \cite{Hawkins1996,Hawkins1995} pairwise generalized Born model is used, with parameters described by Tsui and Case \cite{Tsui2001} (option igb=1 in AMBER 11 input file). The progress coordinates were selected and ``binned" using a $10\times10$ partition of a 2D space. 
A dihedral distance $D=\sqrt{\frac{1}{N}\sum_{i}{d^2_i}} $ $\in  [0,180]$ with respect to a reference set of torsions is used in the first dimension, where $N$ the number of torsional angles considered and $d_i$ is the circular distance between the current value of the $i$-th angle and our reference, i.e., the smaller of the two arclengths along the circumference. This dimension was divided every $14^{\circ}$ from 0 to $126^{\circ}$ and then a final partition covering the space $(126,180])$ 
In the second dimension, a regular RMSD, using only heavy atoms, is measured with respect to an $\alpha$-helical structure. In this case, the space was divided every 0.4\AA\ from 0 to 3.6\AA\ and then a final partition covering the space $[3.6,\infty)$. Values and coordinates for the references used to compute the order parameters are given in the Supporting Information (SI).

The methane molecules were simulated using GROMACS 4.5 software package \cite{gromacs4} with the united-atom GROMOS 45a3 force field \cite{G45A3} and dodecahedral periodic box of TIP3P water molecules \cite{Jorgensen1983} (about 900 water
 molecules in a 34x34x24\AA\ box). The single progress coordinate was the distance $r$ between the two methane molecules, following \cite{Zwier2011}. The coordinate $r \in [0,\infty) \mathrm{\AA}$ 
 was partitioned with a bin spacing of 1\AA\ from 0 to 16\AA\ and a last bin covering the space $r \in [16,\infty) \mathrm{\AA}$.

For the post analysis of methane, different bins were used to demonstrate the flexiblity of the approach.  
The coordinate $r \in [0,\infty) \mathrm{\AA}$ was partitioned so that the first bin is the space $r \in [0,5) \mathrm{\AA}$, then a bin spacing of 2{\AA} were used from 5 to 17, while the last bin covers the space $r \in [17,\infty) \mathrm{\AA}$.

The results shown below include \emph{all} data generated in all trajectories: no transient or relaxation period has been omitted.

\section{Results}
\subsection{Ala4}
\begin{figure}
\begin{center}
   \includegraphics[scale=0.65]{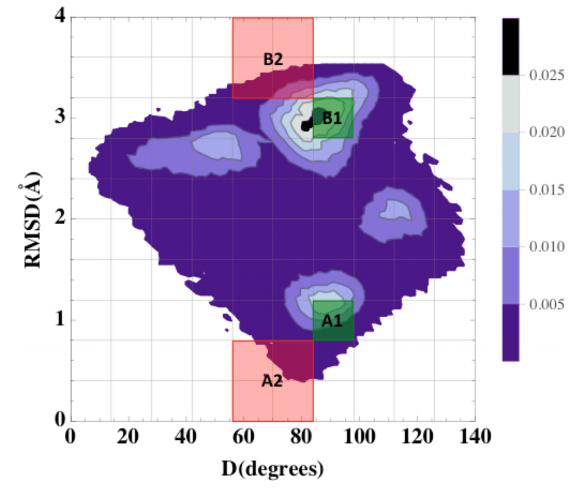}
  \caption{
  \label{fig:ala2Dpmf}
  The Ala4 free energy surface.  The surface is projected onto two coordinates: $D=\sqrt{\frac{1}{N}\sum_{i}{d^2_i}} \in [0,180]$ from one reference structure (see SI) and the RMSD with respect to an ideal $\alpha$-helix. The surface was computed using 3.0 $\mu$s of ordinary ``brute force'' simulation.  The set of states {A1,B1} is highlighted in green, while the second set {A2,B2} is highlighted in red. The grid shows bins that were used both for WE simulation and the post-analysis calculation of observables via the non-Markovian matrix formulation. }
\end{center}
\end{figure}

For Ala4, populations and MFPTs are estimated using WE and compared to independent measurements based on ordinary ``brute force'' (BF) simulation.  Rates are estimated in both directions between the two sets of states {A1,B1} and {A2,B2} shown in Fig.\ \ref{fig:ala2Dpmf} (see SI to visualize representative structures). The second set is less populated and consequently expected to be more difficult to sample. Fig.\ \ref{fig:ala2Dpmf} also shows the bin definitions used in the post-analysis, which were the same as those used during the WE simulation. However, as we shall see in our second system,  we can use any partition of the space for the post analysis.

The data shown below are based on the same total simulation times in BF and WE.  The BF estimates and confidence intervals are based on a single long trajectory of 3.0 $\mu$s where thousands of transitions between states were observed. Five independent WE simulations were run, each employing a total of 3.0 $\mu$s accounting for all the trajectories. The use of independent WE runs permits straightforward error analysis for comparison with BF.

\subsubsection{ Direct estimation of observables via WE }
\begin{figure}
\begin{center}
   \includegraphics[scale=0.55]{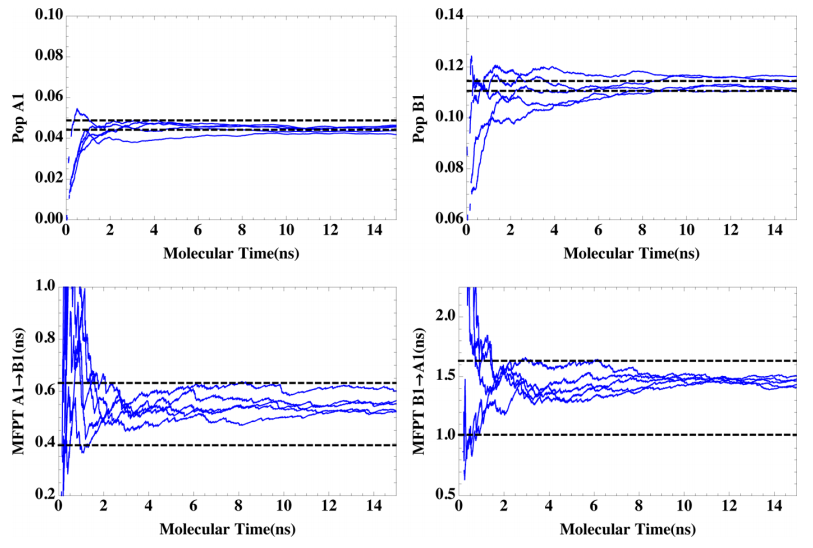}
  \caption{
  \label{fig:states1}
  Direct WE estimates for populations and mean first passage times (MFPTs) for Ala4 states {A1,B1} from Fig. \ref{fig:ala2Dpmf}. Five independent WE runs are shown, each based on 3.0 $\mu$s of total simulation time.  Dashed lines indicate roughly a 95\% confidence interval based on 3.0 $\mu$s of brute force simulation.   Each ns of molecular (single-trajectory) time corresponds to approximately 200 ns of WE simulation including all trajectories in a single run. }
\end{center}
\end{figure}

\begin{figure}
\begin{center}
   \includegraphics[scale=0.55]{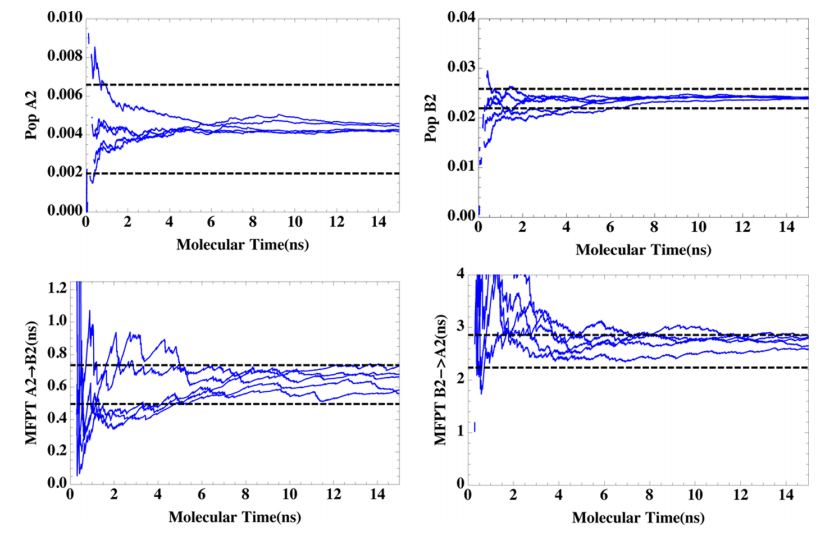}
  \caption{
  \label{fig:states2}
  Direct WE estimates for populations and mean first passage times for Ala4 states {A2,B2} from Fig. \ref{fig:ala2Dpmf}. Five independent WE runs are shown, each based on 3.0 $\mu$s of total simulation time. Dashed lines indicate roughly a 95\% confidence interval based on 3.0 $\mu$s of brute force simulation.   Each ns of molecular time corresponds to approximately 200 ns of WE simulation accounting for all trajectories in a single run. }
\end{center}
\end{figure}

As described above, ``direct'' WE measurements sum trajectory weights for population and flux calculations. Figs. \ref{fig:states1} and \ref{fig:states2}
 show direct estimates for both equilibrium and kinetic quantities for both sets of states.   WE estimates as a function of simulation time are compared to 95\% confidence intervals for BF simulation. 
 
 As with all observables, data from five independent WE simulations is shown. The final/rightmost point from each run is the estimate using all data from the run, and thus is based on a total simulation time equal to that of BF (3 $\mu$s). The spread of the rightmost WE data points therefore can be compared with the BF confidence interval to gauge statistical quality.

The mean values of the direct estimates are in agreement with BF confidence intervals in all cases.  In some cases, the spread of WE estimates is significantly less than that for BF prior to the full extent of WE simulation.  Each ns of ``molecular time'' in Figs.\ \ref{fig:states1} and \ref{fig:states2} (i.e., single-trajectory time) corresponds to approximately 200 nsec of total simulation in a single WE run accounting for all trajectories.  Hence, in some cases, considerably less WE simulation is required for an estimate of the same statistical quality as resulted from the full BF simulation of 3.0 $\mu$s.

\subsubsection{ Non-Markovian matrix analysis }
\begin{figure}
\begin{center}
   \includegraphics[scale=0.6]{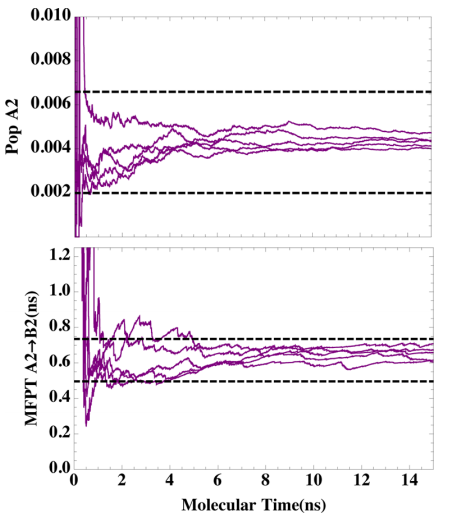}
  \caption{
  \label{fig:matestimate}
  Population of A2 and mean first passage time for Ala4 from A2 to B2, estimated by the non-Markovian matrix analysis of WE data. Dashed lines indicate roughly a 95\% confidence interval from brute force simulation, as in Figs.\ \ref{fig:states1} and \ref{fig:states2}. The states are defined in Fig.\ \ref{fig:ala2Dpmf}. }
\end{center}
\end{figure}

We also show results of the non-Markovian matrix analysis for select observables. 
Fig.\ \ref{fig:matestimate} shows that the non-Markovian analysis yields unbiased estimates of the same equilibrium and non-equilibrium properties calculated with direct estimates. 
(Results for other observables, like the population of A1 and the  A1$\rightarrow$B1 MFPT, not shown, exhibit qualitatively similar agreement.)
The agreement contrasts with a purely Markovian matrix formulation, which does not account for the ``labeling'' described above, which can yield statistically biased estimates for kinetic quantities (see methane results, below). Unbiased matrix-based estimates are important when reweighting is used in WE \cite{Bhatt2010a} as noted in the Discussion. Reweighting was not used in the present study, however.

\subsection{Methane}
In the methane system, WE simulation is used to measure first-passage times based on a range of state definitions. For a complex system, analyzing the sensitivity of the MFPT to state definitions could aid in the definition of states.
 
The MFPT was estimated directly, as well as by both non-Markovian and Markovian matrix analysis.
To assess statistical uncertainty, once again five independent WE simulations were run. The bins used for post-analysis differ from those used in the original WE simulation, as a matter of convenience -  underscoring the flexibility of the approach.

\begin{figure}
\begin{center}
   \includegraphics[scale=0.55]{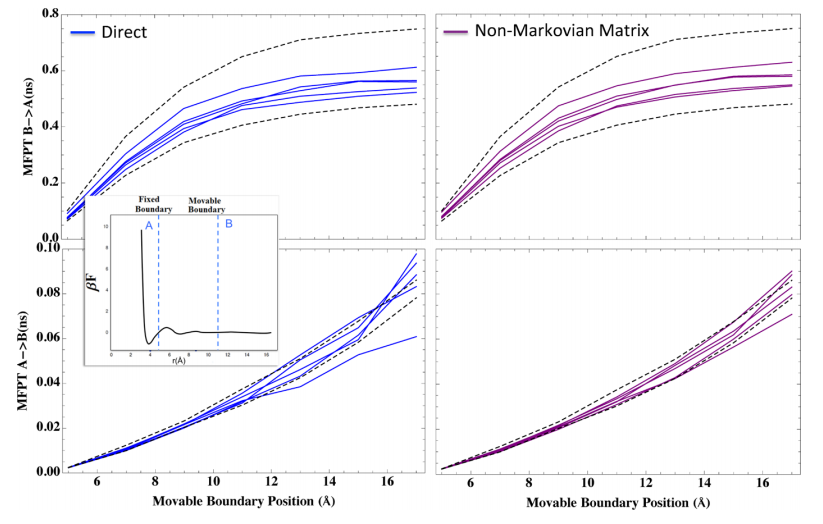}
  \caption{
  \label{fig:mfptvspostion}
  The mean first passage time for methane association (B to A) and dissociation (A to B) measured ``directly'' and from the non-Markovian matrix analysis from WE
simulation as a function of the boundary of state A. The inset displays the PMF along with the definitions of the unbound and bound states, indicated by B and A, respectively. Dashed lines indicate roughly a 95\% confidence interval based on 0.4 $\mu$s of brute force simulation.}
\end{center}
\end{figure}

Fig.\ \ref{fig:mfptvspostion} shows passage times measured as function of the boundary position for the unbound state. The boundary of the bound state A was held fixed at a separation of 5 \AA\ while the definition of the unbound state was varied from 5 to 17 \AA. The passage times were measured in increments of 2 \AA\ and and compared with BF results as shown in Fig.\ \ref{fig:mfptvspostion}. The BF confidence intervals are based on a single long trajectory of 0.4 $\mu$s, the same total simulation time used in each WE simulation. 

Fig.\ \ref{fig:mfptvspostion} shows that both direct and non-Markovian matrix estimates are in agreement with BF confidence intervals.

\begin{figure}
\begin{center}
   \includegraphics[scale=0.55]{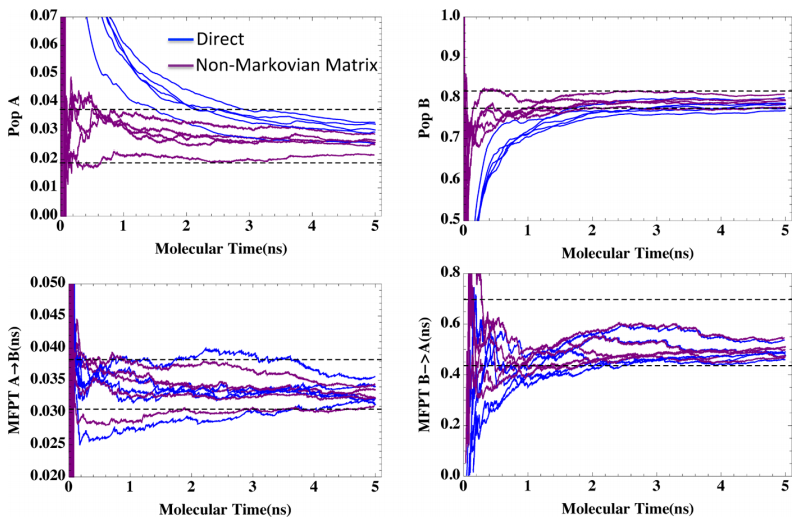}
  \caption{
  \label{fig:directvsmatrix}
Methane association/dissociation observables.
Direct and non-Markovian WE estimates for populations and mean first passage times (MFPTs) are plotted vs.\ molecular time. Five independent WE runs are shown, each based on 0.4 $\mu$s of total simulation time.  Dashed lines indicate roughly a 95\% confidence interval based on 0.4 $\mu$s of brute force simulation. Each ns of molecular time corresponds to approximately 80 ns of WE simulation accounting for all trajectories in a single run. The bound state (A) is defined by distances less than 5 \AA\ and B is defined by distances greater than 11 \AA.}
\end{center}
\end{figure}

For fixed state definitions, Fig.\ \ref{fig:directvsmatrix} shows the evolution of state populations MFPTs, as was done for Ala4.
We fix the the movable boundary position in Fig.\ \ref{fig:mfptvspostion} (inset), defining state B as all configurations with $r>11$\AA.

The performance of the non-Markovian matrix estimates are particularly noteworthy in Fig.\ \ref{fig:directvsmatrix}.
The matrix estimates converge faster than direct estimates to the exact results for the state populations. 
Presumably, this is because the direct approach requires relaxation of the full probability distribution to equilibrium, whereas the matrix approach requires only relaxation of the distribution with each bin (in order to obtain accurate inter-bin rates $\kmunu_{ij}$).

\begin{figure}[h!]
\begin{center}
   \includegraphics[scale=0.55]{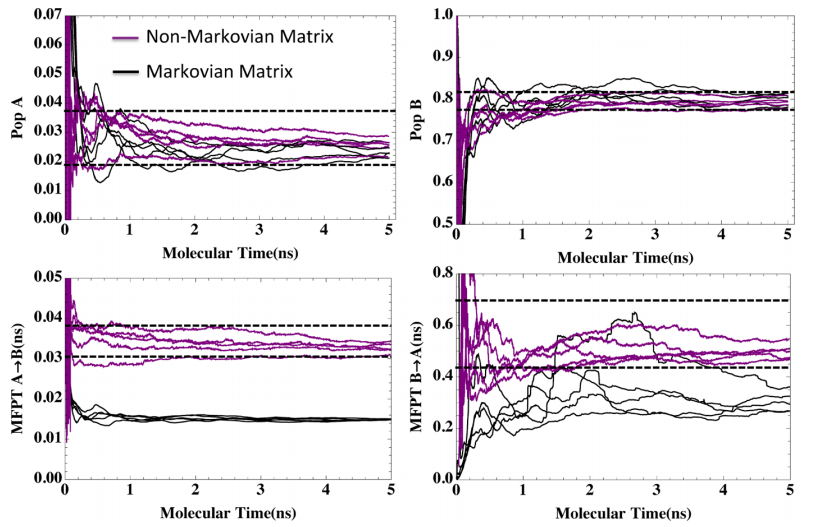}
  \caption{
  \label{fig:colorAndNoColor}
Populations of A($r<5$\AA) and B($r>11$\AA) and MFPTs for the methane system, estimated by the non-Markovian matrix analysis without history information. Dashed lines indicate roughly a 95\% confidence interval from brute force simulation based on 0.4 $\mu$s of total simulation time.}

\end{center}
\end{figure}

In contrast to the unbiased MFPT estimates obtained by both direct and non-Markovian analysis, the Markov analysis can be significantly biased for the MFPT. Fig.\ \ref{fig:colorAndNoColor} shows that applying the Markovian analysis (Sec.\ \ref{sec:nocolorestimation}) leads to MFPT estimates clearly outside the BF confidence interval. Data in the SI shows that the use of a more sophisticated model such as a maximum-likelihood estimator for reversible Markov models \cite{Beauchamp2011} yields similar results and does not correct the bias.

Equilibrium properties, however, can be estimated without bias in a Markovian analysis because history dependence is immaterial. Fig.\ \ref{fig:colorAndNoColor} also illustrates correct (equilibrium) population estimates based on the Markovian analysis.

\section{Discussion}
To our knowledge, this is the first weighted ensemble (WE) study using the original Huber and Kim algorithm \cite{Huber-1996} to simultaneously calculate both equilibrium and non-equilibrium quantities.  The present study estimates observables (populations and MFPTs) based on arbitrary states defined in a post-simulation analysis, permitting the examination of different state definitions and their effects on observables.  Two qualitatively different estimation schemes were examined, including a non-Markovian rate-matrix formulation which shows promise for reducing transient initial-state bias (a bias which is intrinsic to direct estimation of observables based on weights).  
Both schemes, showed substantial efficiency gains for some observables even in the test systems which appear to lack significant energy barriers in their configurational landscapes.  All results were validated using independent ``brute force'' simulations.  Nevertheless, as described below, the present data does point to further challenges likely to be exhibited by larger, more complex systems.

\noindent
\emph{Flexibility in State Choice}\\
\indent One key feature of the WE implementation studied here is the ability to investigate a range of state choices.  As computer simulations tackle systems of growing complexity, it seems increasingly unlikely that states chosen prior to a study will prove physically or biochemically relevant.  Indeed, it is already the case that specialized algorithms are invoked to identify physical states, separated by the slowest timescales, from existing trajectories \cite{Chodera2007,Zhang2010}. With WE simulation, as suggested by our methane data, one can adjust state boundaries to minimize the sensitivity of rates to those boundaries.

A possible concern with post-simulation state construction is the need to store a potentially large set of coordinates to ensure sufficient flexibility in post analysis.  However, modern hardware should be sufficient for most cases of interest.  As an illustration, storage of $\{x, y, z\}$ coordinates for 1,000 heavy atoms in a WE run of 1,000 iterations using 1,000 trajectories would require $\sim$10 GB.

\noindent
\emph{Simultaneous calculation of non-equilibrium and equilibrium observables}\\
\indent The estimation of both equilibrium and kinetic properties from relatively short simulations is an important goal of current methods development, including for WE \cite{Bhatt2012, Darve-chapter}.  Here, we have demonstrated as a ``proof of principle'' that WE simulation can do this efficiently (compared to brute force simulation), without bias, in parallel, and with flexibility in defining states.  Given the relatively fast timescales (nanosecond scale) characterizing the present systems, it is somewhat surprising that WE is better than brute-force simulation for some of the observables and never worse.  Previous studies suggest that WE has the potential for greater efficiency in more complex systems \cite{Zwier2011,Adelman2011,Bhatt2009a}.

\noindent
\emph{Reweighting and the matrix formulation}\\
\indent This study compared estimation of equilibrium and non-equilibrium observables using the original WE algorithm and via post-analysis.  As mentioned in the introduction, the occasional rescaling of weights to match an equilibrium or non-equilibrium steady-state condition \cite{Bhatt2010a} was not used to avoid any potential complications.

Our data clearly show that a standard Markovian analysis of WE simulation is inadequate (Fig. \ref{fig:colorAndNoColor}), since WE bins typically are not Markovian. 
Additional information -- history dependence, as embodied in the $\alpha/\beta$ labeling scheme -- is needed to obtain unbiased results. Inclusion of history information in the matrix analysis means it is intrinsically ``non-Markovian'' regardless of the linear algebra employed.

Future work will incorporate the rate estimation and non-Markovian matrix schemes developed here, as well as possibly the simpler Markovian scheme shown in section \ref{sec:nocolorestimation}.  Our data (Fig.~\ref{fig:directvsmatrix}) suggest these could very successful in bringing a WE simulation closer to a specified steady state.  But it is an open question whether reweighting simulations will prove superior to the type of post-analysis suggested here.  Importantly, data presented here indicate that some rate estimators could lead to biased estimates for populations, which, in turn, would bias a reweighted simulation.

One practical future approach, suggested by the work of Darve, Izaguirre and coworkers \cite{DarveIzaguirre2012}, could be to define preliminary states in advance to aid sampling transitions in both directions, and then to subject the data to the same post analysis performed here to examine additional state definitions besides the initial choices.

\noindent
\emph{Limitations and future work}\\
\indent The present study has not addressed some of the intrinsic limitations of the WE approach, which are the related issues of correlations among trajectories (due to the replication and merging events) and sampling ``orthogonal'' coordinates not divided up by WE bins.  In the systems examined here, there was sufficient sampling in orthogonal dimensions to obtain excellent agreement with brute force results in all cases.  However, significant future effort will be required to address correlations and orthogonal sampling, the latter being a problem common to methods which pre-select coordinates such as multiple-window umbrella sampling  \cite{Haydock-1990,Dickson2011,Dickson2009} and  metadynamics \cite{Vargiu2008,Bussi2006a,Raiteri2006}.

\section{Conclusions}
In this proof-of-principle study, the parallel weighted ensemble (WE) approach has been applied to measure equilibrium and kinetic properties from a single simulation in small but non-trivial molecular systems.  Importantly, populations and rates could be measured for arbitrary states chosen after the simulation.  For all tested observables, unbiased estimates were obtained, as validated by independent brute-force simulations.  In a number of instances, WE was significantly more efficient -- yielding estimates of a given statistical quality in less overall computing time compared to simple simulation, including all trajectories.  In this sense, not only is WE a parallel method, but it can exhibit ``super-linear scaling'', e.g., 100 cores can yield desired information more than 100 times faster than single-core simulation.

We also developed a non-Markovian matrix approach for analyzing WE or brute-force trajectories, capable of yielding unbiased results, sometimes faster than direct estimates of observables from WE.  The non-Markovian formulation also yields simultaneous estimates of equilibrium and non-equilibrium observables based on an arbitrary division of phase space, which is not possible in a standard Markovian analysis.

The approaches tested here will need to be further developed and tested in more complex systems.

\section*{Acknowledgements}
We thank Josh Adelman for insightful discussions, as well as the NSF for support through grants  MCB-0643456, MCB-1119091 and MCB-0845216.

\section*{Supporting Information Available}
Reference coordinates for the order parameters in Ala4, visualization of the states used in Ala4 (A1, A2, B1 and B2) and a comparison of the regular Markov model vs the Maximum Likelihood Estimator for reversible Markov models are shown.
This material is available free of charge via the Internet at http://pubs.acs.org/.

\bibliography{dmz}

\end{document}